\documentclass{article}
\usepackage{frascatiphys}
\usepackage{graphicx}
\begin{document}
\title{ A DEMOCRATIC RESUMMATION PROCEDURE OF SOFT GLUON EMISSION FOR HADRONIC INELASTIC CROSS-SECTIONS AND SURVIVAL PROBABILITIES \footnote{This work was based on collaboration with D. Faundes, A. Grau and O. Shekovtsova.}}
\author{
Giulia Pancheri\\
{\em INFN Frascati National Laboratories, Via E. Fermi 40, Frascati, 00044 Italy}\\
Yogendra N. Srivastava\\
{\em 
Physics 
Department of Physics \& Geology, University of Perugia, Via A. Pascoli, Perugia, 00123, Italy}
}
\maketitle
\baselineskip=10pt
\begin{abstract}
We discuss a model for soft gluon re-summation based on a statistical description of independent emissions during inelastic collisions. The model is applied to estimate Survival Probabilities at the LHC. A comparison with other models and experimental data is presented.
\end{abstract}
\baselineskip=14pt

\section{Introduction}
Survival probabilities at LHC energies are of special interest  when looking for hard scattering events which  need to be selected from the large hadronic background accompanying them. The concept was introduced in \cite{DK}, later defined and discussed in \cite{bjorken}. We recently presented our estimates and discussed them in comparison with other models in \cite{ourSP}. In this contribution, we shall summarize our findings and the particulars of the model we use for calculating the total and the inelastic cross-sections. 

As discussed in \cite{ourSP}, the probability to find events devoid of  hadronic background in the central region can be obtained  in its simplest form  as:
\begin{equation}
\mathcal{S}^2(s)=\int d^2{\bf b} A(b,s) P_{no-hadr-bckg}^{ND}(b,s) 
\label{eq:SP}
\end{equation}
where $P_{no-hadr-bckground}^{ND}(b,s)$ represents the probability of events without activity in the central rapidity region, which can be approximated as the non-diffractive (ND) region of phase space. This is clearly an approximation. However, our aim, as it was in \cite{ourSP},  is to give an order of magnitude estimate of the Survival probabilities, and compare it with other existing predictions. 
The  quantity $A(b,s)$ refers to the normalized distribution of such events in impact parameter space, and   the problem is  to calculate the function $A(b,s)$ appropriate to those events excluded by $P_{no-hadr-bckg}^{ND}(b,s)$.

In the sections to follow, we shall describe our model for these two quantities and  present our phenomenological analysis for $\mathcal{S}^2(s)$.
To estimate survival probabilities following \cite{bjorken} we shall  use the model for the total cross-section we developed in \cite{Corsetti96,Godbole2005}. This model is based on i) single channel eikonal formalism, ii) QCD mini-jets to drive the rise of the total cross-section, iii) soft gluon emission to tame the rise that leads to a high energy behaviour consistent with the Froissart bound. 
\section{Mini-jets vs total cross-sections} 
\begin{figure}[htb]
    \begin{center}
        {\includegraphics[scale=0.6]{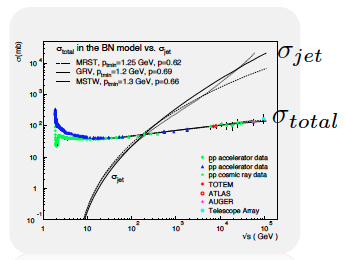}}\hspace{0.5cm}
        {\includegraphics[scale=0.6]{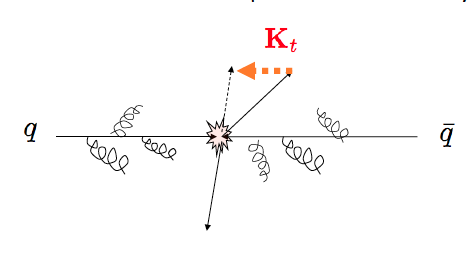}}
        \caption{\it a) The mini-jet proton-proton cross section for inelastic events compared with the total cross-section as a function of c.m. energy b) the soft gluon emission mechanism proposed 
        to tame the fast mini-jet rise.}
\label{fig:minijets-soft}
    \end{center}
\end{figure}

Our suggestion is to extract  the quantities, $A(b,s)$ and $P_{no-hadr-bckg}^{ND}(b,s)$, from  single channel mini-jet models \cite{durand}. We start with the following expressions for the tototal  cross-section: 
\begin{equation}
\sigma_{total}= 2\int d^2 {\bf b} \Im m F_{el}(b,s)=2\int d^2 {\bf b}[1-exp(-\chi_I(b,s))]=\int d^2 {\bf b}[1-exp{(-\bar{n}(b,s)/2)}]
\label{eq:sigtot}
\end{equation}
where the imaginary part of the eikonal function is obtained from the average number of inelastic hadronic collisions. In this approximation, the inelastic total cross-section obtains as 
\begin{equation}
\sigma_{inel}=\int d^2 {\bf b}[1-exp{(-\bar{n}(b,s))}]= \int d^2 {\bf b}[1-exp({-A_{FF}(b,s)\sigma_{soft}(s)-A_{mini-jets}(b,s)\sigma_{mini-jets}(s)})]
\label{eq:siginel}
\end{equation}
with $\sigma_{soft}(s)$  
either a constant or a slowly decreasing function of energy,
and $\sigma_{mini-jets}(s)$ is calculated from perturbative QCD, i.e. using Parton Distribution Functions  (PDFs) DGLAP evolved and folded with  parton-parton cross-sections.
Our mini-jet calculation uses the asymptotic freedom expression for the strong coupling constant and thus implies using a lower cut-off for outgoing partons, $p_{tmin}$, which effectively separates perturbative and non-perturbative collisions. We show in the left panel of Fig. \ref{fig:minijets-soft} 
the behaviour of $\sigma_{mini-jets}(s)$ when calculated for different LO PDFs, and different values of $p_{tmin}$.   The  comparison with the total cross-section  shown in  the same figure, indicates that a mechanism to slow down the excessive growth of the mini-jet cross-section at high energy must be present. In our model, such taming of the mini-jet cross-section is obtained through   the average parton distribution function in impact parameter space  $A_{mini-jets}(b,s)$, for which a distinctive choice based on soft gluon emission processes is made, as we shall describe in the next section.  As for the $b$-distribution of  non-mini-jet events $A_{FF}(b)$,   the present version of the model  is obtained from   the Fourier transform of  the proton e.m. form factor. 

\section{Soft Gluon Re-summation : a democratic pathway through confinement}
\begin{figure}[htb]
    \begin{center}
        {\includegraphics[scale=0.6]{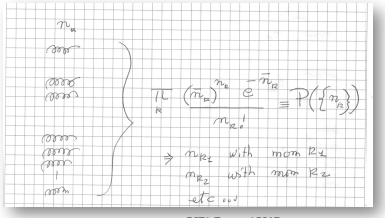}}\hspace{0.5cm}
        \caption{\it  Democratic emission of $n_{\bf k}$ soft gluons of  given momentum $\bf k$  is shown at   left side, with, at right, the overall probability  described by the product of  Poisson distributions.}
\label{fig:democratic}
    \end{center}
\end{figure}

In the right hand panel of Fig. \ref{fig:minijets-soft} we show the mechanism which we propose to be responsible for the taming of the mini-jet effect, soft gluon re-summation (SRG). To tackle SRG, we proceed with the following guiding ideas:
\begin{itemize}
\item if the total cross-section has to follow the limitations of the Froissart bound, hadronic interactions must exhibit  a large distance cut-off,
\item the large distance behaviour (Froissart bound) is  controlled by contributions from  very low momentum gluons, i.e. gluons with  momentum lower than  the pQCD cut-off $\Lambda_{QCD}$,
\item since very soft emitted gluons are not   individually  counted,  only missing energy-momentum is the observed quantity, and  the development of  a formalism for infrared gluons  requires  energy momentum balance to  be enforced  on the soft gluon sea.
\end{itemize}
We propose to use a  semiclassical re-summation procedure, inspired by what was originally proposed in \cite{EPT} for soft photons. This approach   is based on a {\it democratic} treatment,  a term   we shall render more explicit below, and which is represented graphically on  the  left-hand side   in 
Fig.~\ref{fig:democratic}. Let us start with a discrete description of the process of emission.  Additional details about the
soft-gluon re-summation model can be found in our review\cite{Pancheri:2017}.

Let $n_{\bf k}$ be the number of gluons  emitted with  a   given    momentum value $\bf k$. If these gluons are {\it soft}, they are by definition indistinguishable, and independent from the source. 
Hence, the first assumption: these $n_{\bf k}$ gluons, all having exactly the same momentum $k<\Lambda_{QCD}$, are all emitted independently from each other (and from the source).   In analogy  to what  Bloch and Nordsieck demonstrated \cite{BN} for the case of QED,    for each value of the momentum $k$  the number of soft gluons $n_{\bf k}$ is taken as being  distributed according to a Poisson distribution around an average value $\bar{n}_{\bf k}$. The next step in the derivation of the expression we propose, is to consider all possible values of the soft gluon momentum $\bf k$, 
each contributing equally to the final energy momentum imbalance. 
Thus we obtain an overall probability for emission as the product over $\bf k$ of the individual Poisson distributions, i.e. 
\begin{equation}
P(\{n_{\bf k}\})=\Pi_{\bf k}
 \frac{
[\bar{n}_{\bf k}]^{n_{\bf k}}
}{n_{\bf k}!}
e^{
-\bar{n}_{\bf k}
}
\label{eq:nk}
\end{equation}
The next three steps are:\\
1: for each  possible number  of gluons, $n_{\bf k}$, impose energy-momentum conservation, i.e. $K_\mu= \sum_{\bf k} n_{\bf k}k_\mu$,\\
2: considering the distribution in transverse momentum, sum on all the distributions  giving  the observed missing   transverse momentum ${\bf K}_t$, \\
3: exchange the product with the sum,\\
4: take the continuum limit.\\
Explicitly, from 
\begin{equation}
d^2P({\bf K}_t)=
\sum_{n_{\bf k}}
P(
\{
n_{\bf k}
\}) 
d^2{\bf K}_t  \delta^2({\bf K}_t-\sum _{\bf k} {\bf k}_t n_{\bf k})=\sum_{n_{ k}}  \Pi_{\bf k}
 \frac{
[\bar{n}_{\bf k}]^{n_{\bf k}}
}{n_{\bf k}!}
e^{
-\bar{n}_{\bf k}
} d^2{\bf K}_t  \delta^2({\bf K}_t-\sum_{\bf k} {\bf k}_t n_{\bf k})
\end{equation}
and  using the integral representation of the delta-function, one exchanges the sum with
the product obtaining
\begin{equation}
d^2P({\bf K}_t)=\frac{d^2 {\bf K}_t}{(2\pi)^2}
 \int d^2 {\bf b}
  e^{
  -i{\bf K}_t \cdot {\bf b}
  }
   exp\{
   -\sum_{\bf k}{\bar n}_{\bf k}
   [
   1-e^{
   i {\bf k}_t\cdot {\bf b}}
   ]
   \} 
\end{equation}
Going to the continuum, brings
\begin{equation}
d^2P({\bf K}_t)=\frac{d^2 {\bf K}_t }{(2\pi)^2}\int d^2 {\bf b} e^{-i{\bf K}_t \cdot {\bf b}} exp\{-\int d^3{\bar n}_{\bf k}[1-e^{i {\bf k}_t\cdot {\bf b}}]\} \label{eq:d2pk}
\end{equation} 
 Taking then the  Fourier transform of Eq.~(\ref{eq:d2pk}), we obtain the impact parameter distribution 
 as an input  into the eikonal formalism for the inelastic hadronic cross-section, namely
\begin{equation}
A_{mini-jets}\equiv A_{BN}(b,s)={\cal N}(s)e^{-h(b,s)}
\end{equation}
with ${\cal N}(s)$ the normalisation factor, required for dimensional reasons.
Following derivations in our previous publications, we have \cite{Godbole2005}
\begin{equation}
h(b,s)=\frac{8}{3\pi^2}\int_0^{qmax(s)} d^2k_t[1-e^{i
{\bf k}_t\cdot {\bf b}}]
\alpha_s(k_t^2) \frac{\ln (2q_{max}/k_t)}{k_t^2}\label{eq:hbs}
\end{equation}
Of notice are the two limits of integration, the upper limit which is chosen from the kinematics of single gluon emission, and the lower limit, which, in our model,  we put to zero. Thus, a specification for the coupling of soft gluons to their source is needed, since the asymptotic freedom expression of  pQCD cannot be used for $k_t\le \Lambda_{QCD}$.  For such $k_t$ values, we propose a phenomenological ansatz of a singular but integrable behaviour, namely $\alpha_s(k_t^2)\rightarrow (\Lambda_{QCD}/k_t)^{2p}$,  with the condition $1/2<p<1$.  Then the integral in Eq.~(\ref{eq:hbs}) is finite, and the total cross-section is found to behave asymptotically as $\sigma_{tot} \simeq (\ln s)^{1/p}$ \cite{ourfroissart}.

Because our model for re-summation was inspired by the Bloch and Nordsieck theorem,  we refer to it as the {\it BN model}.
\section{The inelastic cross-section and survival probabilities}
\begin{figure}[htb]
    \begin{center}
        {\includegraphics[scale=0.6]{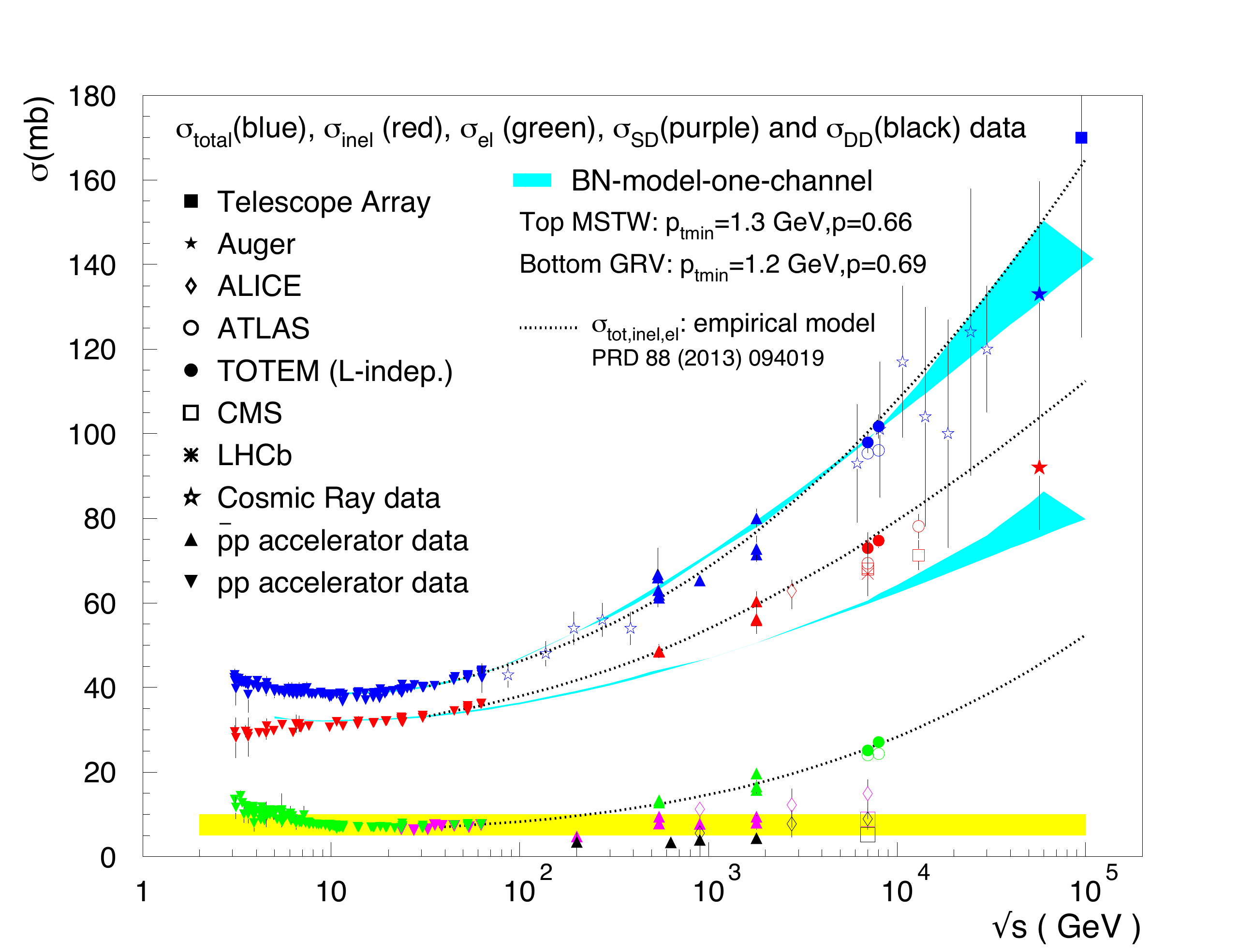}}\hspace{0.5cm}
      \caption{\it  Proton-proton total cross sections, with blue bands corresponding to the BN model expectations, and  dotted lines  are   fits  through an empirical parametrisation \cite{empirical}.}
\label{fig:all}
    \end{center}
\end{figure}

We now apply the above model to describe  proton data  for  the total and inelastic hadronic cross-section in the available energy c.m. range. Applying  Eqs.~ (\ref{eq:sigtot}) and (\ref{eq:siginel}), we 
obtain the curves shown in  Fig.~\ref{fig:all}, with the blue band indicating the uncertainty arising, at very high energies,  from the different low-x- behaviour of the proton PDFs used in the mini-jet calculation. 
 Fig.~ \ref{fig:all} summarises the results obtained with the mini-jet  model described in the previous sections,   and compares them with results from an  empirical model, which  was fashioned after \cite{PB}, and which provides a fit to the total, the elastic and, by subtraction, to the inelastic cross-section.  
  The empirical model is shown by the dotted lines, which are obtained  using an  elastic scattering amplitude parametrised as  \cite{empirical}
\begin{equation}
{\cal A}(s,t)=i [F^2_p(t)\sqrt{A(s)}e^{B(s) t /2}+e^{i\phi(s)}\sqrt{C(s)}e^{D(s)t/2}]
\end{equation}
with $F_p(t)$ the e.m. proton form factor.

 Comparing the fit with the blue band, confirms that the mini-jet model  used here for $\sigma_{inel}$ does non include  diffractive events.  This was discussed in our previous publications \cite{ourSP} where we also noted that    Single Diffractive (SD) events constitute about 10\% of the full inelastic cross-section (approximately indicated  by the yellow band in Fig.~\ref{fig:all}). Their origin can be connected to hadronic products from single hard QCD bremsstrahlung from the quarks in one of colliding protons, but are not  described by  a  single-channel eikonal model, with  only two components in the eikonal, a non perturbative  one, and one from  mini-jets, calculable from semi-hard gluon-gluon scattering.  However, the model we have presented can be used in the calculation of the survival probability when searching for events unaccompanied by  hadronic semi-hard  activity in the central region.
 
\begin{figure}[htb]
    \begin{center}
{\includegraphics[scale=0.4]{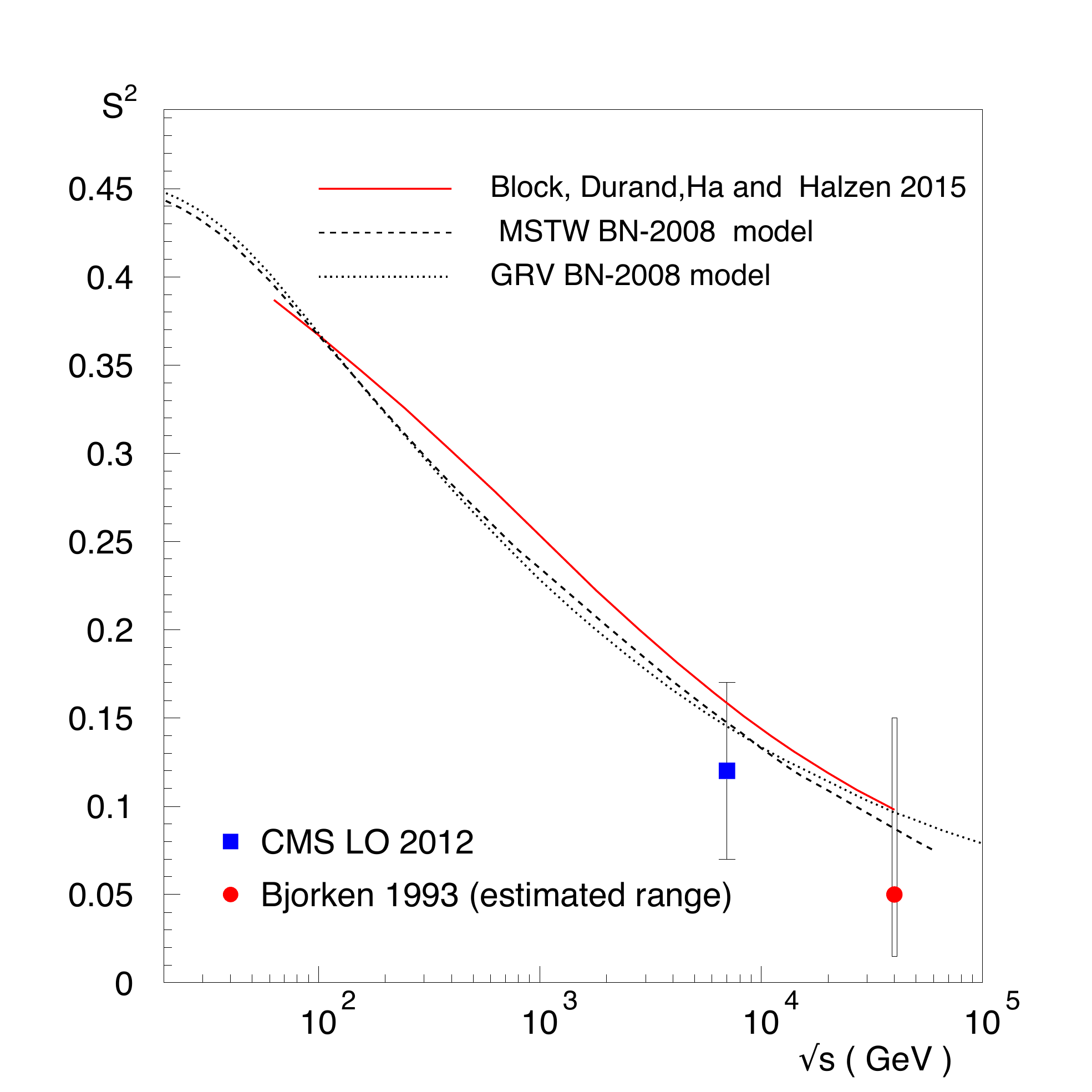}\hspace{0.5cm}\vspace{-0.21cm}
          \includegraphics[scale=0.4]{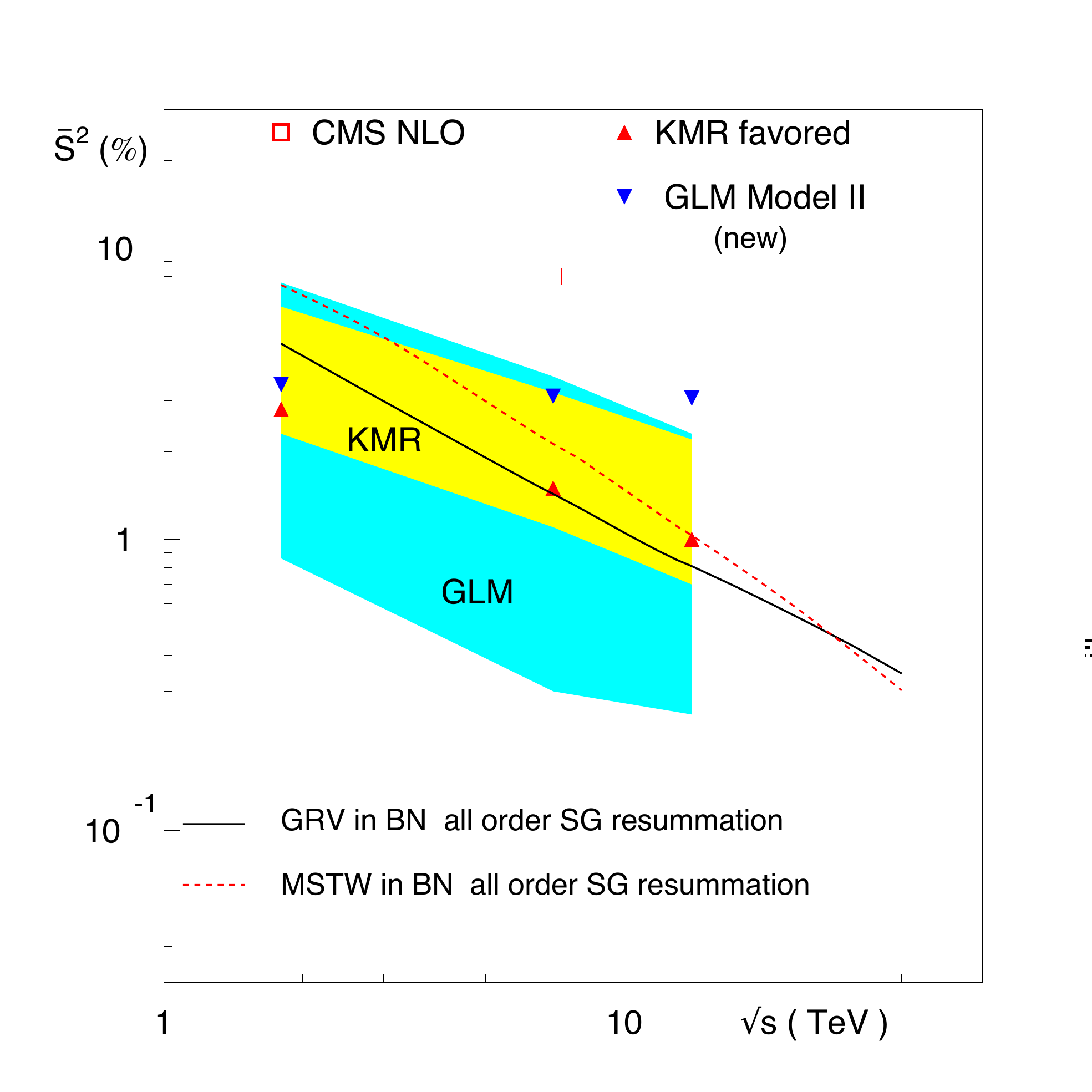}}
     \vspace{1cm}
        \caption{\it 
      Our results for   the survival probability,  when selecting hard events without accompanying hadronic activity in the central region, are compared with other estimates:
        a) when the impact parameter distribution is modelled after the proton e.m. form factor, dotted and dashed lines,  {\it vs.} estimates by  Bjorken \cite{bjorken} and Block, Durand, Ha  and Halzen \cite{BDHH}  (the full red line),  b)  when an  additive  model, described in the text, is used,  compared with  results from  Khoze, Martin and Ryskin   \cite{KMR}, Gotsman, Levin and Maor  from \cite{GLM}. CMS data are from \cite{CMS}, similar results have also been presented by ATLAS \cite{ATLAS}.} 
\label{fig:surv}
    \end{center}
\end{figure}

 In Fig.~\ref{fig:surv} we show  results for the survival probability, estimated using two different  models. For the curves shown at left, we use  Eq.~(\ref{eq:SP}), with  the probability of events with no hadronic background in the central region given as $P_{no-hadr-bckg}^{ND}(b,s) =exp\{-{\bar n}(b,s)\}$, with  the ${\bar n}(b,s)$ function determined through our description of the  inelastic cross-section, as described in the previous section.  As for the impact parameter distribution, we have used $A(b)=A_{FF}(b)$, namely  the Fourier transform of the electromagnetic form factor. This follows previous estimates,  from Bloch, Durand, Ha  and Halzen \cite{BDHH} and our BN model as well (BN-2008 model) \cite{oldSP}. 
 Our  improved proposal \cite{ourSP} is shown in   panel  b) of Fig.~\ref{fig:surv}, where   dotted and full curves correspond to different PDFs in the mini-jet calculation. These curves are obtained using  the results of the previously described  BN model  into an expression for  the survival probability, where  soft and a mini-jet contributions are estimated according to their overall weight,  as follows:
\begin{equation}
\bar{\mathcal{S}}^{2} (s)= \bar{\mathcal{S}}^{2} _{soft}(s)+\bar{\mathcal{S}}^{2}_{mini-jets} (s)
\equiv w_{soft}(s)<| S(b)|^2>_{soft}+w_{mini-jets}(s)<| S(b)|^2>_{mini-jets} 
\label{eq:survsum}
\end{equation}
with
\begin{eqnarray}
<| S(b)|^2>_{soft}=
\int d^2{\bf b} A_{FF}(b,s)e^{-{\bar n}_{soft}(b,s)} \\
<| S(b)|^2>_{mini-jets}= 
\int d^2{\bf b} A_{BN}(b,s)e^{-{\bar n}_{mini-jets}(b,s) }\\
{\bar n}_{soft}(b,s)=A_{FF}(b)\sigma_{soft}(s), \ \ \ \ \  {\bar n}_{mini-jets}(b,s)=A_{BN}(b,s)\sigma_{mini-jets}(s)
\\
w_{soft/mini-jets}(s)\equiv \frac{\sigma_{soft/mini-jets}(s)}{\sigma_{soft}(s)+\sigma_{mini-jets(s)}}
\end{eqnarray}
Our proposed additive model   is    compared in  panel b) of Fig.~\ref{fig:surv}  with other model predictions \cite{BDHH,GLM,KMR}, as well as with  CMS  data for the survival probability associated to diffractive jet production at LHC \cite{CMS}. 

Comparing  results between the two panels and within each figure,  we see  very large differences, of almost one order of magnitude, and also large uncertainties,  between the various estimates. Following our present picture of how mini-jet events populate the central region, we propose     Eq.~(\ref{eq:survsum}) as an  adequate way to develop  a realistic approximation of survival probabilities in the central region.

This contribution  is based on  recent joint work with our collaborators, Agnes Grau, Daniel A. Fagundes and Olga Shekhovotsova. 
YS would like to thank the Department of Physics \& Geology at the University of Perugia for their hospitality.

\end{document}